\documentclass[letterpaper]{article} 
\usepackage{aaai2026}  
\usepackage{times}  
\usepackage{helvet}  
\usepackage{courier}  
\usepackage[hyphens]{url}  
\usepackage{graphicx} 
\def\UrlFont{\rm}  
\usepackage{natbib}  
\usepackage{caption} 
\usepackage{algorithm}
\usepackage{algorithmic}
\usepackage{amsmath}
\usepackage{amssymb}
\usepackage{amsthm}
\newtheorem{remark}{Remark}
\usepackage{stmaryrd}
\usepackage{physics}
\newcommand{\indep}{\perp \!\!\!\! \perp}
\newcommand{\given}{\,|\,}
\newcommand{\E}{\mathbb{E}}
\usepackage{tikz}
\usepackage{xfp} 
\usepackage{pgffor} 
\usepackage{subcaption}
\usepackage{booktabs}
\usepackage{newfloat}
\usepackage{listings}
\DeclareCaptionStyle{ruled}{labelfont=normalfont,labelsep=colon,strut=off} 
\floatstyle{ruled}
\newfloat{listing}{tb}{lst}{}
\floatname{listing}{Listing}
\title{Interpolated Stochastic Interventions Based on Propensity Scores, Target Policies and Treatment-Specific Costs}
\author{
    Johan de Aguas
}
\affiliations{
    Integreat -- Norwegian Centre for Knowledge-driven Machine Learning, Department of Mathematics, University of Oslo \\
    johanmd@uio.no
}

\usepackage{bibentry}

\begin{document}

 \maketitle

\begin{abstract}

We introduce two families of stochastic interventions with discrete treatments that connect causal modeling to cost-sensitive decision making. The interventions arise from a cost-penalized information projection of the independent product of the organic propensity scores and a reference policy, yielding closed-form Boltzmann--Gibbs couplings. The induced marginals define modified stochastic policies that interpolate smoothly, via a tilt parameter, from the organic law or from the reference law toward a product-of-experts limit when all destination costs are strictly positive. The first family recovers and extends incremental propensity score interventions, retaining identification without global positivity. For inference on the expected outcomes after these policies, we derive the efficient influence functions under a nonparametric model and construct one-step estimators. In simulations, the proposed estimators improve stability and robustness to nuisance misspecification relative to plug-in baselines. The framework can operationalize graded scientific hypotheses under realistic constraints. Because inputs are modular, analysts can sweep feasible policy spaces, prototype candidates, and align interventions with budgets and logistics before committing experimental resources.
\end{abstract}

 \begin{links}
     \link{Extended version}{http://arxiv.org/abs/2511.11353}
 \end{links}

\section{Introduction}

Evaluating the causal effects after actions, interventions, and treatment regimes is fundamental to the design and optimization of decision-making systems. Classical causal inference tasks are typically framed via \emph{hard interventions}:  hypothetical scenarios where the exposure is deterministically set to a fixed value for all units in a population \cite{PearlCausality,imbens2015causal}. Although foundational, hard interventions can be overly rigid for many real-world domains, such as healthcare and economic policy, where key information needed to take action is often incomplete, treatment allocation is constrained by limited resources, or when  inference becomes unstable because the data-generating process (DGP) assigns zero probability to certain treatments in specific subpopulations \cite{diaz2013assessing}.

\emph{Stochastic\! /\! soft interventions} offer an expressive alternative, capturing hypothetical scenarios in which the treatment assignment mechanism is altered through probabilistic or functional shifts, typically as a function of observed covariates, while the rest of the DGP is left unchanged \cite{Correa_Bareinboim_2020,CorreaNeurips}. This formulation enables researchers to define, analyze, and estimate causal effects under interventions that are more adaptive, flexible, and aligned with the practical constraints of real-world policy implementation \cite{haneuse2013estimation,sarvet2023longitudinal}.

Among the broad class of stochastic interventions, \emph{incremental propensity score interventions} (IPI) \cite{kennedy2018nonparametric} have gained particular appeal. IPIs tilt the organic propensity score by a tunable parameter, yielding controlled and interpretable odds ratio modifications that allow analysis across a continuum of  intensities \cite{bonvini2023incremental}. They integrate naturally with modern causal workflows and align closely with real-world policy design. Applications include dropout and censoring \cite{kim2021incremental}, time-fixed and time-varying treatments \cite{naimi2021incremental,rudolph2022estimation}, resource-constrained decision making \cite{sarvet2023longitudinal}, interventional mediation \cite{diaz2020causal,hejazi2023nonparametric}, fairness assessment \cite{mcclean2024fair,opacic2025disparity}, and general sensitivity analysis \cite{levis2024stochastic}, among others.

Mechanistically, IPIs differ from other soft interventions such as \emph{general stochastic interventions} (GSI) \cite{Correa_Bareinboim_2020,CorreaNeurips} and \emph{modified treatment policies} (MTP) \cite{haneuse2013estimation,diaz2023nonparametric}, which include \emph{shift intervention policies} (SIP) \cite{sani2020identification}. GSIs allow ample changes to the exposure mechanism, for example, reducing the parent set or introducing auxiliary noise, and they may alter the $\sigma$-algebra of its conditional distribution. MTPs set the counterfactual treatment deterministically as a function of the observed treatment and covariates, with SIPs restricting the rule to depend only on the observed treatment. By contrast, IPIs operate solely by tilting the organic propensity score and preserve the underlying measurable structure of the exposure law  \cite{kennedy2018nonparametric,bonvini2023incremental}. Recently, \emph{generalized policies} (GPs) have been proposed, defined as a marginal of an optimal transport map, matching the distribution induced by a reallocation mechanism conditional on covariates, observed treatment, and auxiliary noise. They have been shown to offer favorable properties for partial identification under latent confounding \cite{levis2024stochastic}.

The expected outcome after an IPI can be identified without the global positivity condition required for hard interventions, provided conditional ignorability holds with a given a backdoor admissible adjustment set. This is important because positivity is often violated in practice, especially with high-dimensional covariates or in longitudinal and dynamic treatment settings \cite{kennedy2018nonparametric,bonvini2023incremental}. When identifiable, the expected outcome after an IPI can be estimated with efficient semiparametric methods that offer favorable statistical properties, including robustness to misspecification and root-$n$ consistency. Common implementations use either \emph{one-step} estimators \cite{bonvini2023incremental} or \emph{targeted minimum loss estimation} (TMLE) \cite{naimi2021incremental}. Inference can be conducted pointwise, at a fixed tilt parameter (for example, a specific odds-ratio shift), or uniformly over a range of values to recover the full intervention response curve \cite{kennedy2018nonparametric}.

\subsection{Motivation and Contribution}

While standard IPIs interpolate between a non-intervention and a hard intervention, many applications call for more nuanced targets. Logistical considerations may prescribe specific treatment shares across arms (for example, treatment 1 to 80\% of units, treatment 2 to 10\%, and no treatment to 10\%), or require a binary exposure to be split 50/50 to preserve fairness constraints. In addition, with multiple treatment arms, deployment costs can vary widely, and finite budgets may limit the overall reach of particular arms.

To address this, we introduce a formulation that yields two families of stochastic interventions indexed by a single tilt parameter. The first family interpolates smoothly, in a cost-aware manner, from the organic law toward its \emph{product-of-experts} (PoE) blend with a pre-specified reference\,/\,target policy, while the second interpolates from the reference policy toward the PoE. The PoE limit is valid when all treatment levels carry strictly positive costs; if some actions are costless, the limiting behavior is determined by the associated zero-cost subset. Both interpolations arise as the marginals of the bivariate distribution that solves a \emph{cost-penalized information projection}. The first family directly generalizes IPIs to incorporate explicit cost structures and target policies and, as in the IPI framework, avoids the global positivity condition for identification. In addition, when the reference distribution is non-degenerate, the induced target marginal itself defines a valid stochastic intervention under an appropriate positivity condition.

For estimation and inference of mean outcomes after these interpolated policies, we derive the efficient influence functions in a nonparametric model and develop one-step estimators that are robust to outcome model misspecification. We further construct asymptotically valid confidence intervals and uniform confidence bands over a specified segment of the interpolation range.

These ideas support the evaluation of hypotheses that depart gradually from the status quo while explicitly accounting for costs, enabling analysts to prototype candidate policies and align interventions with budgetary and logistical constraints before committing experimental resources, with relevance in domains such as healthcare and economics.

\section{Preliminaries}

\subsection{Hard Interventions}

Let $A\in\mathcal{A}$ be a discrete point-exposure,  $Y\in\mathbb{R}$ a continuous outcome, $W\in\mathcal{W}$ a vector of pre-exposure covariates, and $\pi(a\given w):=\mathbb{P}(A=a\given W=w)$ the propensity score of treatment option $a\in\mathcal{A}$. Potential outcomes $Y^a$ encode unit-level counterfactuals after a hard intervention $\operatorname{do}(A=a)$ \cite{PearlCausality}. Under counterfactual consistency, positivity $\pi(a\given w)\in(0,1), \forall w\in\mathcal{W}$, and conditional ignorability\,/\,backdoor admissibility of $W$, one can identify the expected outcome after intervention $\operatorname{do}(A=a)$ from observational data via the $g$-computation\,/\,backdoor formula, as $ \mathbb{E}\left[Y^a \right] = \mathbb{E}\left[Y\given\operatorname{do}(A=a) \right] = \mathbb{E}_W\left\{ Q(W,a) \right\}$, with $Q(w,a)=\mathbb{E}[Y\given W=w,A=a]$.

\subsection{Incremental Propensity Score Interventions (IPI)}

For a binary point-exposure $A$, an IPI can be defined as a parametrized family of stochastic interventions that replace the organic propensity score $\pi(1\given w)$ by:
\begin{equation}
    \widetilde{\pi}_\delta(1\given w):=\frac{e^\delta\pi(1\given w)}{e^\delta\pi(1\given w)+\pi(0\given w)},\quad \delta\in\mathbb{R},
\end{equation}

\noindent for all $w\in\mathcal{W}$, and where $\delta$ is a user-specified tilt parameter that governs the degree of deviation from the organic treatment assignment mechanism. Notably, $\delta$ corresponds to the log-odds ratio between the modified and the organic propensity scores, providing an interpretable parameterization of intervention intensity \cite{kennedy2018nonparametric}.

IPIs provide smooth interpolations between a non-intervention and a hard intervention. Specifically, when $\delta=0$, the intervention leaves the propensity score unchanged, i.e., $\widetilde{\pi}_\delta(a\given w)={\pi}(a\given w)$. As $\delta\to\infty$, the modified mechanism approaches a deterministic assignment to treatment value $a=1$, i.e., $\widetilde{\pi}_\delta(a\given w)\to\mathbb{I}(a=1)$; conversely, as $\delta\to-\infty$, the intervention converges to always assigning $a=0$, i.e., $\widetilde{\pi}_\delta(a\given w)\to\mathbb{I}(a=0)$.

The expected outcome after a stochastic intervention, such as an IPI with modified exposure mechanism $A\sim\widetilde{\pi}_\delta$, can be denoted using notation $\mathbb{E}[Y^{\widetilde{\pi}_\delta}]$. Under conditional ignorability\,/\,backdoor admissibility of $W$, it is identified by:
\begin{equation}
    \mathbb{E}[Y^{\widetilde{\pi}_\delta}] = \sum_{a\in\mathcal{A}}\mathbb{E}_W\left\{\widetilde{\pi}_\delta(a\given W)\, Q(W,a)\right\},
\end{equation}

Notably, unlike the case of a hard intervention, this causal query does not require a global positivity condition \cite{kennedy2018nonparametric,bonvini2023incremental}.

\section{Cost-Penalized I-Projection}

While exponentially tilted distributions have been characterized as standard I-projections under a mean constraint \cite{levis2024stochastic}, and IPIs have been used to model interventions under limited resource and treatment costs \cite{sarvet2023longitudinal}, a broader connection between IPIs and bivariate I-projections that accommodate cost structures and target distributions has not been described. In this work, we provide that characterization.

Given two input probability measures $\pi$ (\emph{source}) and $\nu$ (\emph{target}) over an action set $\mathcal{A}$, a summable\,/\,integrable cost function on pairs $c:\mathcal{A}^2 \!\to[0,\infty)$, and a penalization parameter $\delta\geq 0$, we define the \emph{cost-penalized I-projection} (CPIP) of the independent product $\pi\otimes\nu$  as the joint distribution $\gamma\in\mathcal{M}_+^1(\mathcal{A}^2)$ (in the class of distributions over the Cartesian product $\mathcal{A}^2=\{(A',A'') : A',A''\in\mathcal{A}\}$) that solves:
\begin{equation}\label{eq:ROTnew}
\inf_{\gamma\in\mathcal{M}^1_+(\mathcal{A}^2) } 
 \mathbb{D}_{\operatorname{KL}}(\gamma\given\pi\otimes\nu) +\delta\,\mathbb{E}_{\gamma}\left\{c(A',A'')\right\},
\end{equation}

\noindent where $\mathbb{D}_{\operatorname{KL}}$ represents the Kullback--Leibler divergence. 

This problem is closely related to \emph{unconstrained} or \emph{limiting-case} variants of entropic optimal transport and Schrödinger bridge problems \cite{leonard2014survey, relaxedOT, chizat2018scaling, peyre2019computational}. Yet, while entropic optimal transport problems typically require iterative solvers such as the Sinkhorn--Knopp algorithm, the CPIP problem admits a closed-form solution thanks to the strong convexity and smoothness of its objective. Its unique minimizer is given by the Boltzmann--Gibbs kernel:
\begin{equation}\label{eq:solution}
    \gamma^\star_\delta(a',a'') \propto \pi(a')\,\nu(a'')\,e^{-\delta c(a',a'')}.
\end{equation}

\subsection{Tilted Marginal Distributions}

Let $A$ be a categorical point-exposure variable with domain $\mathcal{A}=\{\alpha_1,\dots,\alpha_{K}\}$, where the $K$ treatment options may potentially include a placebo or null-treatment. Let $\nu(\cdot\given w)$ be a target distribution over $\mathcal{A}$ and let $\pi(\cdot\given w)$ be the propensity scores given pre-exposure covariate profile $W=w$. Let $c(a',a'')$ be the cost of reallocating a unit from treatment $A=a'$ to treatment $A=a''$, which does not depend on the profile $w$. Then, for all $a\in\mathcal{A}$ and $w\in\mathcal{W}$,
\begin{align}
    \pi^\star_\delta(a\given w) &:= \frac{\pi(a\given w)\sum_{a''\in\mathcal{A}}\nu(a''\given w)\,e^{-\delta c(a,a'')}}{\sum_{a',a''\in\mathcal{A}}\pi(a'\given w)\,\nu(a''\given w)\,e^{-\delta c(a',a'')}},\\
    \nu^\star_\delta(a\given w) &:= \frac{\nu(a\given w)\sum_{a'\in\mathcal{A}}\pi(a'\given w)\,e^{-\delta c(a',a)}}{\sum_{a',a''\in\mathcal{A}}\pi(a'\given w)\,\nu(a''\given w)\,e^{-\delta c(a',a'')}}
\end{align}

\noindent are the marginal distributions of the CPIP solution in \eqref{eq:solution} with source $\pi$ and  target $\nu$. We refer to these as the \emph{tilted source\,/\,target marginal distributions}. Derivations are provided in the technical appendix A1.

\begin{remark}\label{rem:IPI}
Let:
\begin{enumerate}
\item $A\in\mathcal{A}=\{0,1\}$ be a binary point-exposure,
\item the target marginal $\nu$ be the degenerate distribution that always assigns treatment, $\nu(a\given w)=\mathbb{I}(a=1)$,
\item $c(a',a'')=\mathbb{I}(a'\neq a'')$ be the Hamming cost.
\end{enumerate}

Then, the tilted source marginal of the CPIP solution with penalization parameter $\delta$ coincides with an IPI with tilt parameter $\delta$, and thus $\widetilde{\pi}_\delta(1\given w)=\pi^\star_\delta(1\given w)$ for all $w\in\mathcal{W}$.
\end{remark}

We provide a proof in the technical appendix A2.

This remark provides an interpretation of IPIs in the binary exposure setting with $\delta>0$. Suppose the target intervention assigns treatment to all units, and the reallocation cost is 1 for any switch between $A=0$ and $A=1$  while maintaining the same treatment status has no cost. Then an IPI with tilt parameter $\delta$ coincides with the first marginal of a joint distribution that minimizes the expected reallocation cost, subject to a penalty (with coefficient $1/\delta)$ on its divergence from the independent product distribution $\pi(a' \given w)\, \mathbb{I}(a'' = 1)$. Such product places zero mass on pairs $(a', 0)$ and assigns mass to pairs $(a', 1)$ equivalent to the propensity score $\pi(a'\given w)$.

The CPIP objective is strictly convex and well posed for $\delta\geq 0$, yielding a unique solution. For $\delta<0$, the joint distribution $\gamma^\star_\delta$ in \eqref{eq:solution} is not the optimizer of the CPIP program; however, when the action set $\mathcal{A}$ is finite and costs are bounded, the associated Boltzmann--Gibbs kernel is summable, so $\gamma^\star_\delta$ remains normalizable and produces smooth and closed-form \emph{parametric extensions} for $\pi^\star_\delta$ and $\nu^\star_\delta$. In this extrapolative regime, the induced tilted marginals shift mass toward higher-cost and repulsive pairings and are pushed away from the independence law $\pi\otimes \nu$, favoring more adversarial couplings. For continuous spaces with potentially unbounded costs, additional integrability or tail conditions are required for $\delta<0$ to deliver well posed stochastic policies.

Notably, within the setting of remark \ref{rem:IPI}, the tilted target marginal coincides with the input target distribution, so $\nu^\star_\delta(a\given w)=\mathbb{I}(a=1)$ for all $\delta\geq 0$ and all $w\in\mathcal{W}$. Consequently, it does not yield a differentiated stochastic policy. In the next remark, we introduce a generalization of IPIs that accommodates treatment-specific costs and a flexible target policy. Although we state the results for an unconditional reference\,/\,target $\nu(a)$, the extension to a conditional target $\nu(a\given w)$ is straightforward.

\begin{remark}\label{rem:new}
Let:
\begin{enumerate}
\item $A\in\mathcal{A}=\{\alpha_1,\dots,\alpha_{K}\}$ be a categorical point-exposure variable with $K$ treatment options,
\item The target marginal $\nu$ be any valid probability distribution over $\mathcal{A}$,
\item the reallocation cost from $A=\alpha_j$ to $A=\alpha_k\neq \alpha_j$ be a value that is specific for the received treatment $\alpha_k$ and constant over profiles $W=w$, i.e., $c(\alpha_j,\alpha_k)=c(\alpha_k)\,\mathbb{I}(\alpha_j\neq \alpha_k)$, with $0\leq c(a)<\infty$ for all $a\in\mathcal{A}$.
\end{enumerate}

Then, for each $w\in\mathcal{W}$, the tilted marginals of the CPIP solution with parameter $\delta$ are:
\begin{align}\label{eq:transf1}
    \pi^\star_\delta(a\given w) &:= \frac{(\zeta_\delta+\xi_\delta(a))\,\pi(a\given w)}{\sum_{a'\in\mathcal{A}}(\zeta_\delta+\xi_\delta(a'))\,\pi(a'\given w)},\\ \label{eq:transf2}
    \nu^\star_\delta(a\given w) &:= \frac{\nu(a)-\xi_\delta(a)(1-\pi(a\given w))}{\sum_{a'\in\mathcal{A}}(\zeta_\delta+\xi_\delta(a'))\,\pi(a'\given w)},
\end{align}


where:
\begin{align}
    \xi_\delta(a) &:=\nu(a)\left(1-e^{-\delta c(a)}\right),\\
    \zeta_\delta &:=\sum_{a'\in\mathcal{A}}\nu(a')\,e^{-\delta c(a')}.
\end{align}        
\end{remark}

We provide a derivation in the technical appendix A3.

Note that $\pi^\star_0(a\given w)=\pi(a\given w)$ and $\nu^\star_0(a\given w)=\nu(a)$ for all $a\in\mathcal{A}$ and $w\in\mathcal{W}$. In other words, setting $\delta=0$ results in no modification of the input distributions. Furthermore, denote $\mathcal{A}_0=\{a\in\mathcal{A}: c(a)=0\}$, $\mathcal{A}_+=\{a\in\mathcal{A}: c(a)>0\}$, $\nu^\dagger(a):=\nu(a)\,\mathbb{I}(a\in\mathcal{A}_+)+\sum_{a\in\mathcal{A}_0}\nu(a)$ and $\pi^\dagger(a\given w):=\pi(a\given w)+(1-\pi(a\given w))\,\mathbb{I}(a\in\mathcal{A}_0)$. Then, in the limit $\delta\to\infty$, one obtains:
\begin{align}
    \pi^\star_\infty(a\given w) &=\frac{\pi(a\given w)\,\nu^\dagger(a)}{\sum_{a'\in\mathcal{A}} \pi(a'\given w)\,\nu^\dagger(a') },\\ \nu^\star_\infty(a\given w) &=\frac{\pi^\dagger(a\given w)\,\nu(a)}{\sum_{a'\in\mathcal{A}} \pi^\dagger(a'\given w)\,\nu(a') }.
\end{align}

When all treatment costs are positive, both reduce to the \emph{product of experts} (PoE) distribution $\operatorname{PoE}(a)\propto\pi(a\given w)\,\nu(a)$  \citep{hinton1999products}. This PoE law can be interpreted as a \emph{consensus} distributional blend of inputs $\pi,\nu$. 

We argue that the tilted target distribution $\nu^\star_\delta$ can also be used to define stochastic policies, particularly when the target $\nu$ is non-degenerate. For example, suppose the ideal policy prescribes treatment 1 to 80\% of units, treatment 2 to 10\%, and no treatment to the remaining 10\%. Then, in the case with positive costs, the tilted target marginal $\nu^\star_\delta$ provides a smooth interpolation between this ideal allocation ($\delta = 0$) and the PoE law ($\delta \to \infty$). 

\begin{figure}[t]
    \centering

    \begin{subfigure}[t]{0.22\textwidth}
        \centering
        \includegraphics[page=1,scale=0.875]{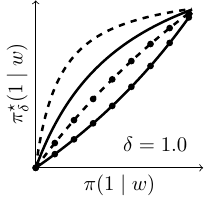}
        \caption{}
    \end{subfigure}
    \begin{subfigure}[t]{0.22\textwidth}
        \centering
        \includegraphics[page=2,scale=0.875]{Tikz_2.pdf}
        \caption{}
    \end{subfigure}

    \begin{subfigure}[t]{0.22\textwidth}
        \centering
        \includegraphics[page=3,scale=0.875]{Tikz_2.pdf}
        \caption{}
    \end{subfigure}
    \begin{subfigure}[t]{0.22\textwidth}
        \centering
        \includegraphics[page=4,scale=0.875]{Tikz_2.pdf}
        \caption{}
    \end{subfigure}

    \caption{Tilted source distributions $\pi^\star_\delta$ in panels (a) and (b), and tilted target distributions $\nu^\star_\delta$ in panels (c) and (d), for a binary exposure. Curves show pointwise transformation of the propensity score $\pi(1\given w)$ for $\delta\in\{1.0,2.5\}$. Target configuration is encoded by dot pattern: $\nu=(0,1)$ is non-dotted and $\nu=(0.7,0.3)$ is dotted. Cost structure is encoded by line style: solid for $c=(1,1)$ and dashed for $c=(0,2)$. Vector components are ordered as $(A=0, A=1)$.}
    \label{fig:2x2}
\end{figure}

Figure \ref{fig:2x2} presents the tilted marginal distributions for a binary exposure under varying configurations of the tilt parameter $\delta$, cost functions $c$, and target distribution $\nu$. 

\newcommand{\fval}[6]{%
  \fpeval{
    ( (#1*exp(-#5*#3) + #2*exp(-#5*#4)) + #2*(1 - exp(-#5*#4)) )*#6 /
    (
      ( (#1*exp(-#5*#3) + #2*exp(-#5*#4)) + #2*(1 - exp(-#5*#4)) )*#6 +
      ( (#1*exp(-#5*#3) + #2*exp(-#5*#4)) + #1*(1 - exp(-#5*#3)) )*(1 - #6)
    )
  }%
}

\newcommand{\gval}[6]{%
  \fpeval{
    ( #2 - #2*(1 - exp(-#5*#4))*(1-#6) ) /
    (
      ( (#1*exp(-#5*#3) + #2*exp(-#5*#4)) + #2*(1 - exp(-#5*#4)) )*#6 +
      ( (#1*exp(-#5*#3) + #2*exp(-#5*#4)) + #1*(1 - exp(-#5*#3)) )*(1 - #6)
    )
  }
}

\newcommand{\coordsf}[6]{%
  \def\coords{}%
  \foreach \x in {0,0.02,...,1.0} {
    \xdef\coords{\coords (\x,\fval{#1}{#2}{#3}{#4}{#5}{\x})}
  }
}

\newcommand{\coordsg}[6]{%
  \def\coords{}%
  \foreach \x in {0,0.02,...,1.0} {
    \xdef\coords{\coords (\x,\gval{#1}{#2}{#3}{#4}{#5}{\x})}
  }
}

\newcommand{\plotf}[6]{%
  \coordsf{#1}{#2}{#3}{#4}{#5}{}%
  \draw[#6, very thick] plot[smooth] coordinates {\coords};
}

\newcommand{\plotg}[6]{%
  \coordsg{#1}{#2}{#3}{#4}{#5}{}%
  \draw[#6, very thick] plot[smooth] coordinates {\coords};
}

\subsection{Pushforward Distribution}

The CPIP criterion optimizes over joint laws $\gamma(\,\cdot,\cdot\given w)$ without imposing or penalizing any marginal constraint. Therefore the solution $\gamma^\star_\delta$ in \eqref{eq:solution} should not be viewed as a transport plan that maps $\pi$ to $\nu$. In particular, the tilted marginal $\nu^\star_\delta(\cdot\given w)$ is simply the second marginal of $\gamma^\star_\delta$ and does not represent the distribution obtained by pushing $\pi(\cdot\given w)$ through $\gamma^\star_\delta$. If one wants a transport-style interpretation, one can modify the objective by enforcing the marginals to equal or remain close to the inputs via penalties, as in relaxed optimal transport \cite{relaxedOT,chizat2018scaling,chizat2018unbalanced}. Such formulations, however, typically sacrifice the closed-form solution. As an operational alternative within the CPIP construction, we can define a pushforward of the original source law by using the Markov kernel induced by $\gamma^\star_\delta$:
\begin{equation}
    \nu^{\sharp}_\delta(a\given w) \;:=\; \sum_{a'\in\mathcal{A}}\frac{\gamma^\star_\delta(a',a\given w)}{\pi^\star_\delta(a'\given w)}\, \pi(a'\given w),
\end{equation}

This is well defined provided $\pi^\star_\delta(a'\given w)>0$ when $\pi(a'\given w)>0$. For profile-independent destination costs $c(a',a''\given w)=c(a'')\,\mathbb{I}(a'\neq a'')$, the kernel does not depend on $a'$, and the pushforward reduces to an exponential tilting of $\nu$, i.e., 
$\nu^{\sharp}_\delta(a\given w)\propto \nu(a)\,e^{-\delta c(a)}$.
A systematic comparison between the CPIP marginal, marginal-penalized OT variants, and the pushforward is left for future work.

\section{Identification}

Let $\mu^S_\delta \equiv \mathbb{E}\!\left[Y^{\pi^\star_\delta}\right]$ and $\mu^T_\delta \equiv \mathbb{E}\!\left[Y^{\nu^\star_\delta}\right]$ denote the expected outcomes after the stochastic interventions associated with the tilted marginals in remark \ref{rem:new}. These functionals are identified from observational data under conditional ignorability\,/\,backdoor admissibility.

Let $W=\operatorname{pa}(A;\mathcal{G})$ denote the causal parents of $A$ in the assumed causal graph $\mathcal{G}$ of the system. In typical structures without latent confounding between $A$ and $Y$, the set $W$ is backdoor admissible \cite{PearlCausality}. However, $W$ may include instrumental variables or predictors of $A$ that are unrelated to $Y$; such variables do not induce bias but can reduce precision \cite{cinelliControls}. This motivates the use of a possibly smaller backdoor admissible set $Z\subseteq W$ such that $Y\indep_d (W\setminus Z)\given Z,A$ in $\mathcal{G}$. If such set exists, then:
\begin{align}\label{eq:functional1}
    \mu^S_\delta &=  \sum_{a\in\mathcal{A}}\mathbb{E}_W\left\{\pi^\star_\delta(a\given W)\, Q(Z,a)\right\},\\ \label{eq:functional2}
   \mu^T_\delta &=  \sum_{a\in\mathcal{A}}\mathbb{E}_W\left\{\nu^\star_\delta(a\given W)\, Q(Z,a)\right\}, 
\end{align}

\noindent where $Q(z,a)=\mathbb{E}[Y\given Z=z,A=a]$ and $\mathbb{E}_W$ denotes expectation with respect to the marginal distribution of $W$ (equivalently, of $Z$ since $Z\subseteq W$).

For \eqref{eq:functional2} to be well defined and estimable, a global positivity condition is required: 
\begin{equation}
    \sup_{a\in\mathcal{A}} \frac{\nu^\star_\delta(a\given W)}{\pi(a\given W)}<\infty,\ \ P_W\text{-almost surely}.
\end{equation}
 
A sufficient condition is uniform overlap on the support of $\nu$, namely the existence of constants $\{\varepsilon_a>0\}_{a\in\mathcal{A}}$ such that $\pi(a\given W)\ge \varepsilon_a$ $P_W$-almost surely for all $a$ with $\nu(a)>0$. By contrast, the corresponding global positivity condition for $\mu^S_\delta$ holds automatically, since $\pi^\star_\delta(a\given W)$ is a bounded multiplicative tilt of $\pi(a\given W)$. Hence, no additional overlap is needed for $\mu^S_\delta$, in accordance with the corresponding property of IPIs \cite{kennedy2018nonparametric}.

\section{Semiparametric Efficient Estimation}

To develop estimators of the expected outcomes, we assume that the cost function $c$ and the target distribution $\nu$ are known and fixed; they are treated as design inputs rather than estimated from data. Plug-in estimators for $\mu^S_\delta$ and $\mu^T_\delta$ are then obtained by estimating $\pi$ and $Q$, applying the transformations in \eqref{eq:transf1} and \eqref{eq:transf2} to construct the tilted marginals, substituting these into the inner product in \eqref{eq:functional1} and \eqref{eq:functional2}, and averaging over i.i.d. samples.

Recent work has leveraged semiparametric efficient methods to estimate causal queries after stochastic interventions \cite{diaz2012population,duong2021stochastic,bonvini2023incremental,diaz2023nonparametric}. Semiparametric efficient approaches are prized for accommodating data-adaptive learning while attaining optimal asymptotics under flexible DGP assumptions \cite{MLDR}. Many estimators enjoy double or multiple robustness, remaining consistent despite some nuisance misspecification \cite{DRbookVariance}. Core frameworks include Newton–Raphson one-step corrections \cite{Pfanzagl1982,bickel1998,robins1995analysis,robins1997comment}, targeted minimum loss estimation (TMLE) \cite{TMLEbook1,TMLEbook2}, and debiased machine learning (DML) \cite{DML}. These methods leverage the \emph{efficient influence function} (EIF) of the smooth functional defining the parameter \cite{hines2022demystifying}.

\begin{remark}\label{rem:last}
Let $\mathcal{S}_\delta[P]$ denote the functional that evaluates expression \eqref{eq:functional1} at an arbitrary distribution $P$. Similarly, let $\mathcal{T}_\delta[P]$ represent the corresponding $P$-functional from expression  \eqref{eq:functional2}. Let $O_i=(W_i,A_i,Y_i)$ denote a sample drawn from the true DGP $P^\circ $. Under a nonparametric model, suitable smoothness and regularity conditions, the \emph{uncentered} EIF of $\mathcal{S}_\delta[P]$ at $P^\circ $, evaluated at point $O_i$, exists and is:
\begin{align}\label{eqEIF1}
& D_{\delta}^{{S}}(O_i) = D_{\delta}^{{S},1}(O_i) + D_{\delta}^{{S},2}(O_i),\\ \notag
& D_{\delta}^{{S},1}(O_i) = \frac{\pi^\star_\delta(A_i\given W_i)}{\pi(A_i\given W_i)}\left[Y_i-\sum_{a\in\mathcal{A}} \pi^\star_\delta(a\given W_i)\,Q(Z_i,a) \right]\\ \notag
& D_{\delta}^{{S},2}(O_i) = \sum_{a\in\mathcal{A}} \pi^\star_\delta(a\given W_i)\,Q(Z_i,a).
\end{align}

Analogously, under a nonparametric model and typical smoothness and regularity conditions, the \emph{uncentered} EIF of $\mathcal{T}_\delta[P]$ at $P^\circ $, evaluated at point $O_i$, exists and is given by:
\begin{align}\label{eqEIF2} 
& D_{\delta}^{{T}}(O_i) = D_{\delta}^{{T},1}(O_i) + D_{\delta}^{{T},2}(O_i) + D_{\delta}^{{T},3}(O_i),\\ \notag
& D_{\delta}^{{T},1}(O_i) = \frac{\nu^\star_\delta(A_i\given W_i)}{\pi(A_i\given W_i)}\left[Y_i-Q(Z_i,A_i) \right]\\ \notag
& D_{\delta}^{{T},2}(O_i) = \left[2-\frac{\pi^\star_\delta(A_i\given W_i)}{\pi(A_i\given W_i)} \right]\sum_{a\in\mathcal{A}} \nu^\star_\delta(a\given W_i)\,Q(Z_i,a) \\ \notag
& D_{\delta}^{{T},3}(O_i) = \frac{\pi^\star_\delta(A_i\given W_i)}{\pi(A_i\given W_i)}\varrho_\delta(A_i)\,Q(Z_i,A_i)  \\ \notag
&\qquad\qquad\quad -\sum_{a\in\mathcal{A}} \pi^\star_\delta(a\given W_i)\,\varrho_\delta(a)\,Q(Z_i,a),\\ \notag
&\text{ where } \varrho_\delta(a) := \frac{\xi_\delta(a)}{\zeta_\delta+\xi_\delta(a)}.
\end{align}
\end{remark}

For binary exposure $A$ and target distribution $\nu(a)=\mathbb{I}(a=1)$, it is straightforward to verify that $D_{\delta}^{{S}}(O_i)$ coincides with the uncentered EIF for the expected outcome after an IPI, as derived in prior work \cite{kennedy2018nonparametric}.

Note that $\varrho_0(a)= 0$ and $\pi^\star_0(a\given w)=\pi(a\given w)$ for all $a\in\mathcal{A},w\in\mathcal{W}$. Hence, when $\delta=0$ and $\nu(a)=\mathbb{I}(a=a')$, we recover the standard uncentered EIF for the expected outcome after the hard intervention $\operatorname{do}(A=a')$.

Consequently, Newton--Raphson one-step estimators of the expected outcomes in \eqref{eq:functional1} and \eqref{eq:functional2} are given by:
\begin{equation}
    \widehat{\mu}^S_\delta = \frac{1}{n}\sum_{i=1}^n \widehat{D}^{S}_\delta(O_i)\quad \text{and}\quad   \widehat{\mu}^T_\delta = \frac{1}{n}\sum_{i=1}^n \widehat{D}^{T}_\delta(O_i),
\end{equation}

\noindent using sample $\{O_i\}_{i=1}^n\overset{iid}{\sim} P^\circ $. Here $\widehat{D}^{S}_\delta(\cdot)$ and $\widehat{D}^{T}_\delta(\cdot)$ are formed by substituting the nuisance estimates $\widehat{\pi}$ and $\widehat{Q}$ into  \eqref{eqEIF1} and  \eqref{eqEIF2}, respectively. These estimators remove the first-order bias in the von Mises expansion of the associated functionals around the true distribution \cite{Pfanzagl1982,bickel1998}, are robust to outcome-model misspecification through $\widehat{Q}$, and are asymptotically linear under standard regularity conditions provided the propensity score estimator satisfies a convergence rate of $\norm{\widehat{\pi}-\pi}_{L^2(P)}=o_{P^\circ }
(n^{-1/4})$, or faster.

Derivation and proof are provided in appendix A4 and A5.

\subsection{Inference}

To relax technical conditions for weak convergence, such as requiring nuisance estimators to belong to a Donsker class, we employ sample splitting and cross-fitting. In this approach, nuisance functions are estimated on one subset of the data and then plugged into the influence function evaluated on a disjoint subset; final estimates average results across folds \cite{DML}. 

Let $\widehat{\sigma}^S_\delta$ and $\widehat{\sigma}^T_\delta$ be consistent estimators of ${\operatorname{var}}\left(D_{\delta}^{{S}}(O)\right)^{1/2}$ and ${\operatorname{var}}\left(D_{\delta}^{{T}}(O)\right)^{1/2}$. Estimation and uncertainty quantification proceed as follows:
\begin{enumerate}
  \item Split the sample into $K$ folds. On each training fold $k$, fit $\widehat{\pi}_k$ and $\widehat{Q}_k$ using data-adaptive learners.
  \item For a grid $G$ of $\delta$ values, construct $\widehat{\pi}^{\star}_{\delta,k}$ and $\widehat{\nu}^{\star}_{\delta,k}$ via \eqref{eq:transf1}–\eqref{eq:transf2}.
  \item Evaluate $\widehat{D}^{S}_{\delta,k}$ and $\widehat{D}^{T}_{\delta,k}$ on the corresponding held-out fold, aggregate across units and folds, and form one-step estimators $\widehat{\mu}^S_\delta$ and $\widehat{\mu}^T_\delta$.
  \item \textbf{Uniform confidence bands}: For $b\in\{1,\dots, B\}$, draw i.i.d. multipliers $\{\chi_i^{(b)}\}_{i=1}^n\overset{iid}{\sim} N(0,1)$, and compute
  \begin{equation}
  \zeta^{S}_b=\sup_{\delta\in G}\left|\frac{1}{\sqrt{n}}\sum_{i=1}^n \chi_i^{(b)}(\widehat D^S_\delta(O_i)-\widehat{\mu}^S_\delta)/\widehat{\sigma}^S_\delta\right|, 
  \end{equation}
  and save as $\xi^S$ the 95\% quantile of $\{\zeta^{S}_b\}_{b=1}^B$ for band construction around $\widehat{\mu}_\delta^S$. Analogously, construct $\xi^T$ using $\widehat{D}_\delta^T$ and $\widehat{\sigma}^T_\delta$.
    \item Report final point estimates and intervals:
    \begin{equation}
    \widehat{\mu}^S_\delta \mp \xi^S\, \frac{\widehat{\sigma}^S_\delta}{\sqrt{n}}\quad\text{ and }\quad  \widehat{\mu}^T_\delta \mp \xi^T\,  \frac{\widehat{\sigma}^T_\delta}{\sqrt{n}}.
\end{equation}
\end{enumerate}

For \emph{pointwise} inference at a fixed $\delta$, take $\xi^S=\xi^T=1.96$ to obtain 95\% Wald-type intervals. For \emph{uniform} bands over $\delta\in G$, use the multiplier-bootstrap critical values $\xi^S$ and $\xi^T$ as above. Since the uncentered EIF, viewed as a function of $\delta$, is Lipschitz continuous in $\delta$ under our construction, the scaled estimation error process converges weakly in $\ell^\infty(G)$ to a tight Gaussian process, validating the uniform bands \cite{chernozhukov2013gaussian,kennedy2018nonparametric}.

\section{Simulations}

We conducted an evaluation task for the proposed estimators using repeated simulations with finite synthetic data. The employed DGP is adapted from \citet{kang2007demystifying} and \citet{kennedy2018nonparametric}, with bespoke modifications to introduce a three-leveled categorical exposure and to increase the noise in the system, thereby aligning the signal-to-noise ratio with amounts commonly observed in social science and observational clinical data. The DGP is given by: 
\begin{align*}
    W &\overset{iid}{\sim} N(\vec{0},I_4),\\
    \eta_1(W) &= \exp(-2W_1 + W_2 - 0.5W_3 - 0.25W_4),\\
    \eta_2(W) &= \exp(-W_1 + 0.25W_2 + 2W_3 + 0.5W_4),\\
    \pi(\alpha_k\given W) &= \eta_k(W)/\left[\eta_1(W)+\eta_2(W)+1 \right],\ j\in\{1,2\},\\
    \pi(\alpha_3\given W) &= 1- \pi(\alpha_1\given W)-\pi(\alpha_2\given W),\\
     A\given W &\overset{iid}{\sim}\operatorname{Cat}_3\left( \pi(\alpha_1\given W),\pi(\alpha_2\given W),\pi(\alpha_3\given W)\right),\\
     q(W) &= 2W_1+W_2+W_3+W_4,\\
     Q(W,A) &= \begin{cases} 10-8.7\, q(W) & \text{ if } A=a_1\\ 40+17.4\, q(W) & \text{ if } A=a_2\\ 50+26.1\, q(W) & \text{ if } A=a_3 \end{cases},\\
     Y &= Q(W,A) + \varepsilon, \text{ where } \varepsilon\overset{iid}{\sim} N(0,50).
\end{align*}

\begin{table}[t]
\small
\centering
\begin{tabular}{lcrrrr}
\toprule
\multicolumn{6}{l}{Setup  1: $c =(2.0,1.0,1.0)$, $\nu =(0.4,0.4,0.2)$}  \\
           &  & \multicolumn{2}{c}{$\widehat{\mu}^S_\delta$ (under $\widehat{\pi}_\delta^\star$)} & \multicolumn{2}{c}{$\widehat{\mu}^T_\delta$ (under $\widehat{\nu}_\delta^\star$)} \\
\cmidrule(lr){3-4} \cmidrule(lr){5-6}
   Estim.        &  Misspec.      & iBias & iRMSE & iBias & iRMSE \\
\midrule
plug-in     & -- & 0.40 & 2.36 & 3.03 & 5.69 \\
plug-in     & $Q$ & 1.99 & 3.21 & 15.22 & 15.43 \\
plug-in     & $\pi$ & 14.00 & 14.57 & 7.92 & 9.48 \\
one-step    & -- & \textbf{0.02} & 2.20 & \textbf{0.57} & 8.44 \\
one-step    & $Q$ & \textbf{0.03} & \textbf{2.20} & \textbf{0.35} & \textbf{13.89} \\
one-step    & $\pi$ & \textbf{0.39} & \textbf{2.32} & \textbf{3.98} & \textbf{5.86} \\
\midrule
\multicolumn{6}{l}{Setup  2: $c =(1.0,0.5,2.0)$, $\nu =(0.5,0.3,0.2)$}  \\
           &  & \multicolumn{2}{c}{$\widehat{\mu}^S_\delta$ (under $\widehat{\pi}_\delta^\star$)} & \multicolumn{2}{c}{$\widehat{\mu}^T_\delta$ (under $\widehat{\nu}_\delta^\star$)} \\
\cmidrule(lr){3-4} \cmidrule(lr){5-6}
   Estim.        &  Misspec.      & iBias & iRMSE & iBias & iRMSE \\
\midrule
plug-in     & -- & 0.21 & 2.20 & 3.43 & 5.66 \\
plug-in     & $Q$ & 0.54 & 2.25 & 17.61 & 17.80 \\
plug-in     & $\pi$ & 14.67 & 15.14 & 3.08 & 5.59 \\
one-step    & -- & \textbf{0.02} & 2.14 & \textbf{0.38} & 7.38 \\
one-step    & $Q$ & \textbf{0.02} & 2.14 & \textbf{0.50} & \textbf{11.24} \\
one-step    & $\pi$ & \textbf{0.36} & \textbf{2.23} & \textbf{2.66} & \textbf{4.82} \\
\midrule
\multicolumn{6}{l}{Setup  3: $c =(1.0,1.0,2.0)$, $\nu =(0.0,0.2,0.8)$}  \\
           &  & \multicolumn{2}{c}{$\widehat{\mu}^S_\delta$ (under $\widehat{\pi}_\delta^\star$)} & \multicolumn{2}{c}{$\widehat{\mu}^T_\delta$ (under $\widehat{\nu}_\delta^\star$)} \\
\cmidrule(lr){3-4} \cmidrule(lr){5-6}
   Estim.        &  Misspec.      & iBias & iRMSE & iBias & iRMSE \\
\midrule
plug-in     & -- & 1.20 & 3.04 & 9.46 & 11.82 \\
plug-in     & $Q$ & 3.82 & 4.77 & 27.41 & 27.76  \\
plug-in     & $\pi$ & 13.32 & 14.12 & 10.28 & 12.68  \\
one-step    & -- & \textbf{0.04} & \textbf{2.40} & \textbf{0.75} & \textbf{8.86} \\
one-step    & $Q$ & \textbf{0.06} & \textbf{2.39} & \textbf{0.49} & \textbf{16.07} \\
one-step    & $\pi$ & \textbf{1.15} & \textbf{2.83} & \textbf{5.17} & \textbf{7.50}  \\
\bottomrule
\end{tabular}
\caption{Integrated bias (iBias) and root mean squared error (iRMSE) for the estimated expected outcome after stochastic interventions $\pi^\star_\delta$ and $\nu^\star_\delta$. Results are averaged over 200 simulations and shown for two estimators (plug-in and one-step), across three model specifications: (i) correctly specified, (ii) misspecified outcome regression $Q$, and (iii) misspecified propensity score $\pi$; and under three combinations of cost functions $c$ and target distributions $\nu$.}
\label{tab:bias-rmse}
\end{table}

For the exposure model class, we use multinomial logistic regression with a linear predictor, and for the outcome we employ multivariate adaptive regression splines (MARS) with extra linear predictors $W$. These model classes are correctly specified in the sense that they contain the true propensity score $\pi$ and outcome regression function $Q$, respectively. Following the approach of \citet{kang2007demystifying}, we introduce ad hoc misspecification in $\pi$ and $Q$ by using the same model classes but replacing the original covariates $W\in\mathbb{R}^4$ with a nonlinear transformation $X(W)\in\mathbb{R}^3$, which also constitutes a valid adjustment set, defined as:
\begin{align}
X_1 &= 10 + W_2 / (1 + \exp(W_1)),\\ X_2 &= (0.6 + W_1W_3 / 25)^3,\\
  X_3 &= (W_2 + W_4 + 20)^2.
\end{align}

We compare the performance of two estimators for the expected outcome after stochastic interventions: the plug-in estimator and the one-step bias-corrected estimator introduced in the previous section. This comparison is carried out for both stochastic intervention strategies $\pi^\star_\delta$ and $\nu^\star_\delta$ across various model misspecification scenarios, cost structures, and target distributions. Estimator performance is evaluated using integrated bias (iBias) and integrated root mean squared error (iRMSE), standard metrics in functional estimation. These measures are computed by averaging the absolute value of $\delta$-specific bias and RMSE results over 100 equally spaced values of $\delta$ in the interval $[-2,2]$.

Table \ref{tab:bias-rmse} reports results for $n=1000$, averaged over $200$ Monte Carlo repetitions. Across all three setups, the proposed one-step estimators consistently outperform the corresponding plug-in estimators. Under correct specification, the one-step estimators attain essentially almost-zero bias and achieve lower or comparable RMSE, with the gains most pronounced after the policy $\pi^\star_\delta$. Under $\nu^\star_\delta$, the reduction in bias can come at the cost of a modest increase in RMSE when the target distribution $\nu$ is diffuse rather than concentrated on a single action. The robustness advantage is clearest when the outcome regression $\widehat{Q}$ is misspecified: the one-step estimators retain low bias and RMSE throughout, whereas the plug-in estimators incur substantial deterioration in both measures, especially after $\nu^\star_\delta$. When the propensity score model $\widehat{\pi}$ is misspecified, consistency is no longer theoretically guaranteed, yet the one-step correction still delivers lower bias and RMSE than plug-in across settings. Overall, $\widehat{\mu}^T_\delta$ is more sensitive to misspecification than $\widehat{\mu}^S_\delta$, but the one-step correction systematically attenuates this sensitivity, underscoring its practical value in finite samples under imperfect nuisance modeling.

\section{Application Case}

\begin{figure}[t]
    \centering

    \begin{subfigure}[t]{0.4\textwidth}
        \centering
        \includegraphics[page=1,scale=0.875]{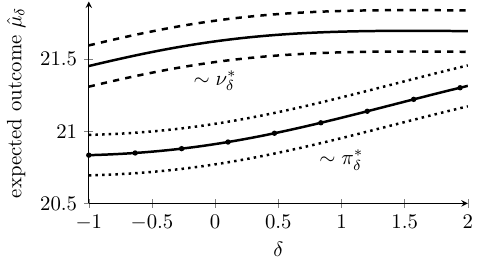}
        \caption{$c=(0.0,1.0,2.0)$ and $\nu=(0.00,0.75,0.25)$}
    \end{subfigure}
    \begin{subfigure}[t]{0.4\textwidth}
        \centering
        \includegraphics[page=2,scale=0.875]{Tikz_1.pdf}
        \caption{$c=(0.5,0.5,2.0)$ and $\nu=(0.00,0.50,0.50)$}
    \end{subfigure}

    \caption{Estimated expected outcome after stochastic interventions {$\pi_\delta^\star$} and {$\nu_\delta^\star$} in the application case with 3-levels exposure: (1) no treatment, (2) low dose, and (3) high dose treatment. Broken lines represent uniform confidence bands.}
    \label{fig:2x1}
\end{figure}

We estimate the expected outcomes after various stochastic intervention strategies involving pharmacological treatment with stimulants for ADHD on children's academic achievement.  We focus on numeracy test scores in grade 8th among Norwegian children diagnosed with ADHD. Drawing on linked data from national registries, we compile comprehensive records on medication histories and academic outcomes for all children born between 2000 and 2007 in Norway who were diagnosed with ADHD and eligible for the national tests up to 2021, excluding those with severe conditions, resulting in a sample of 8,609 kids. Information at the student, family, and school levels is integrated from prescription databases, patient registries, and official health, education and demographic statistics, including medical consultations and diagnostic histories. To estimate  \emph{defined daily doses} (DDD) from prescription records, we apply the PRE2DUP algorithm \cite{tanskanen2015prescription} to classify stimulant treatment into three categories: untreated (47\%), \emph{low dose} (21\%) and \emph{high dose} (32\%).

Although stimulant medication is well-documented to reduce core ADHD symptoms \cite{Cortese2018-qd}, its effects on educational outcomes are more modest \cite{faraone2021world, Storebo2015dz}. Evidence points to only small improvements in standardized test performance \cite{pelham2022effect, JANGMO2019423} and limited long-term gains in academic achievement among Norwegian children \cite{Varnet}. 

 Given that methylphenidate is the most widely prescribed stimulant, and that adverse effects such as sleep disturbances and weight loss are commonly reported \cite{graham2008adverse}, the overall cost of treatment may encompass both direct healthcare expenditures and costs associated with managing adverse effects.  To account for this, we evaluate two cost and target configurations, both assigning a higher cost to high-dose treatment: (\emph{a}) a target in which 75\% of individuals receive low-dose treatment and 25\% receive high-dose treatment, and (\emph{b}) a balanced target with a 50--50\% allocation between low and high doses.

Figure \ref{fig:2x1} displays the estimated expected outcomes after source-tilted  ($\pi^\star_\delta$) and target-tilted ($\nu^\star_\delta$) stochastic interventions across the range $\delta \in [-1, 2]$, with 95\% uniform confidence bands. In both scenarios, expected outcomes increase gradually with $\delta$, indicating a small benefit, of less than one test score point, as the interventions shifts the propensity score toward the PoE distribution. Across most of the range of $\delta$, the target-tilted policy $\nu^\star_\delta$ achieves consistently higher expected outcomes than the source-tilted policy $\pi^\star_\delta$, indicating that, for this case, shifting the target distribution dominates shifting the source. Results also indicate that target distributions placing greater emphasis on low-dose treatment are associated with better academic achievement outcomes.

\section{Conclusions}

We propose a cost-aware family of stochastic interventions for discrete treatments that generalizes incremental propensity score interventions and connects causal modeling to cost-sensitive decisions. Modeling a cost-penalized I-projection of the independent product of organic law and reference\,/\,target policy yields closed-form Boltzmann–Gibbs couplings whose marginals, via a single tilt $\delta$, interpolate from either of the input distributions toward a product-of-experts limit when destination costs are strictly positive. We derive efficient influence functions under a nonparametric model and construct one-step estimators with uniform bands over $\delta$, improving stability and misspecification robustness relative to plug-in baselines.

These policies enable graded scientific hypotheses under realistic constraints. Because $\delta$ is continuous and costs $c$ and targets $\nu$ are modular, analysts can prototype and evaluate policies for prospective studies, turning observational registries into pre-experimental testbeds. Explicit costs clarify prioritization, aligning interventions with budgets and logistics while quantifying trade-offs. Clinician-informed targets integrate naturally, ensuring policies reflect expert priors and empirical regularities. Overall, this framework links identification and estimation from observational data to resource-aware experimental design when hard interventions are infeasible.

\section{Acknowledgments}

This work was supported by the Research Council of Norway via two grants with project numbers 302899 (\emph{Effects of pharmacological treatment and special education on school performance in children with ADHD}) and 332645  (\emph{Integreat -- Norwegian centre for knowledge-driven machine learning}). Data curation for the application case was conducted by Dr.\ Guido Biele at the Norwegian Institute of Public Health. The study was approved by the Regional Committees for Medical and Healthcare Research Ethics (REK approval no.\ 96604).

\bibliography{aaai2026}

\clearpage
\appendix
\onecolumn

\makeatletter
\newcommand{\LeftSection}[2][]{%
  \refstepcounter{section}%
  \addcontentsline{toc}{section}{\protect\numberline{\thesection}#2}%
  \par\bigskip
  {\Large\bfseries\raggedright \thesection\quad #2\par}%
  \medskip
  \if\relax\detokenize{#1}\relax\else\label{#1}\fi
}
\newcommand{\LeftSectionStar}[1]{%
  \par\bigskip
  {\Large\bfseries\raggedright #1\par}%
  \medskip
}
\makeatother

\section*{\LARGE{Technical Appendix}}

\LeftSectionStar{A1 -- CPIP and tilted marginals}

Consider cost-penalized I-projection (CPIP) of the independent product of input distributions $\pi$ and $\nu$ over discrete sample space $\mathcal{A}$, formulated as:
\begin{equation*}
    \inf_{\gamma\in\mathcal{M}^1_+(\mathcal{A}^2) } \sum_{a',a''\in\mathcal{A}} c(a',a'')\,\gamma(a',a'')+\frac{1}{\delta}\mathbb{D}_{\operatorname{KL}}(\gamma\given\pi\otimes\nu),
\end{equation*}

\noindent where $\delta>0$, $\mathcal{M}^1_+(\mathcal{A}^2)$ denotes the set of probability measures over sample space $\mathcal{A}^2$, and $c:\mathcal{A}^2\to\mathbb{R}$ is a nonnegative and integrable cost function. Owing to the strict convexity of the objective, the unique minimizer satisfies the KKT conditions obtained from the Lagrangian:
\begin{equation*}
\mathcal{L}(\gamma,\lambda) = \sum_{a',a''\in\mathcal{A}} \left[ c(a',a'') + \frac{1}{\delta} \log\left( \frac{\gamma(a',a'')}{\pi(a')\,\nu(a'')} \right) \right] \gamma(a',a'') - \lambda \left[ \sum_{a',a''\in\mathcal{A}} \gamma(a',a'') - 1 \right],
\end{equation*}

\noindent which admits a unique closed-form solution. The first-order condition is given by:
\begin{align*}
    & \dv{\mathcal{L}}{\gamma(a',a'')} = c(a',a'')+\frac{1}{\delta}\left[\log\left( \frac{\gamma (a',a'') }{\pi(a')\, \nu(a'')\ } \right) +1\right]-\lambda =0,\\
    \Rightarrow &\ \log\left( \frac{\gamma (a',a'') }{\pi(a')\, \nu(a'')\ } \right) = \delta [\lambda -c(a',a'')]-1,\\
    \Rightarrow &\ \gamma (a',a'') = \pi(a')\, \nu(a'')\, e^{-\delta c(a',a'')}\, e^{\delta\lambda -1 }\propto \pi(a')\, \nu(a'')\, e^{-\delta c(a',a'')}.
\end{align*}

Now suppose $\pi\equiv \pi(\cdot\given w)$ is a conditional distribution, and the cost function $c$ does not depend on the covariate profile $W=w$. Then, for each profile $w$, the solution plan becomes:
\begin{equation*}
    \gamma^\star_\delta(a',a''\given w) = \frac{\pi(a'\given w)\, \nu(a'')\, e^{-\delta c(a',a'')}}{\sum_{a',a''\in\mathcal{A}}\pi(a'\given w)\, \nu(a'')\, e^{-\delta c(a',a'')}}.
\end{equation*}

The marginals of $\gamma^\star(a',a''\given w)$ are then straightforward to compute:
\begin{equation*}
    \pi^\star_\delta(a\given w) = \frac{\pi(a\given w)\sum_{a''\in\mathcal{A}}\nu(a'')\,e^{-\delta c(a,a'')}}{\sum_{a',a''\in\mathcal{A}}\pi(a'\given w)\,\nu(a'')\,e^{-\delta c(a',a'')}}\quad\text{and}\quad
    \nu^\star_\delta(a\given w) = \frac{\nu(a)\sum_{a'\in\mathcal{A}}\pi(a'\given w)\,e^{-\delta c(a',a)}}{\sum_{a',a''\in\mathcal{A}}\pi(a'\given w)\,\nu(a'')\,e^{-\delta c(a',a'')}}.
\end{equation*}

\LeftSectionStar{A2 -- Remark 1: IPIs as special case}

Let $A\in\{0,1\}$ be a binary point-exposure. Let the target marginal $\nu$ be the degenerate distribution that always assigns treatment, $\nu(a)=\mathbb{I}(a=1)$, and let $c(a',a'')=\mathbb{I}(a'\neq a'')$ be the Hamming cost. Then, the tilted source marginal of the CPIP solution with regularization parameter $\delta$ is:

\begin{align*}
    \pi^\star_\delta(a\given w) &= \frac{\pi(a\given w)\sum_{a''\in\mathcal{A}}\mathbb{I}(a''=1)\,e^{-\delta \mathbb{I}(a\neq a'')}}{\sum_{a',a''\in\mathcal{A}}\pi(a'\given w)\,\mathbb{I}(a''=1)\,e^{-\delta \mathbb{I}(a'\neq a'')}}= \frac{\pi(a\given w)\,e^{-\delta \mathbb{I}(a\neq 1)}}{\sum_{a'\in\mathcal{A}}\pi(a'\given w)\,e^{-\delta \mathbb{I}(a'\neq 1)}},\\
    \Rightarrow \pi^\star_\delta(1\given w) &= \frac{\pi(1\given w)\,e^{-\delta \cdot 0}}{\pi(1\given w)\,e^{-\delta \cdot 0}+\pi(0\given w)\,e^{-\delta \cdot 1}} = \frac{\pi(1\given w)\,e^{\delta}}{\pi(1\given w)\,e^{\delta}+\pi(0\given w)},
\end{align*}

\noindent which coincides with an IPI with tilt parameter $\delta$, and thus $\pi^\star_\delta(1\given w)=\widetilde{\pi}_\delta(1\given w)$ for all $w\in\mathcal{W}$.

\LeftSectionStar{A3 -- Remark 2: Tilted marginals under treatment-specific costs}

Let:
\begin{itemize}
\item $A\in\mathcal{A}=\{\alpha_1,\dots,\alpha_{K}\}$ be a categorical point-exposure variable  with $K$ treatment options,
\item The target marginal $\nu$ be any valid probability distribution over $\mathcal{A}$,
\item The reallocation cost from $A=\alpha_j$ to $A=\alpha_k\neq \alpha_j$ be a value that is specific for the received treatment $\alpha_k$ and constant over profiles $W=w$, i.e., $c(\alpha_j,\alpha_k)=c(\alpha_k)\,\mathbb{I}(\alpha_j\neq \alpha_k)$, with $0\leq c(a)<\infty$ for all $a\in\mathcal{A}$.
\end{itemize}

Then, the tilted source marginal of the CPIP solution with parameter $\delta$ corresponds to:
\begin{align*}
    \pi^\star_\delta(a\given w) &= \frac{\pi(a\given w)\sum_{a''\in\mathcal{A}}\nu(a'')\,e^{-\delta c(a'')\,\mathbb{I}(a\neq a'')}}{\sum_{a',a''\in\mathcal{A}}\pi(a'\given w)\,\nu(a'')\,e^{-\delta c(a'')\,\mathbb{I}(a'\neq a'')}} =  \frac{\pi(a\given w)\left[\nu(a)+\sum_{a''\neq a}\nu(a'')\,e^{-\delta c(a'')}  \right]}{\sum_{a'\in\mathcal{A}} \pi(a'\given w)\left[\nu(a')+\sum_{a''\neq a'}\nu(a'')\,e^{-\delta c(a'')}  \right] },\\
    &=  \frac{\pi(a\given w)\left[\nu(a)+\sum_{a''\in \mathcal{A}}\nu(a'')\,e^{-\delta c(a'')} -\nu(a)\,e^{-\delta c(a)} \right]}{\sum_{a'\in\mathcal{A}} \pi(a'\given w)\left[\nu(a')+\sum_{a''\in\mathcal{A}}\nu(a'')\,e^{-\delta c(a'')} -\nu(a')\,e^{-\delta c(a')} \right] }=  \frac{\pi(a\given w)\left(\zeta_\delta + \xi_\delta(a) \right)}{\sum_{a'\in\mathcal{A}} \pi(a'\given w)\left(\zeta_\delta + \xi_\delta(a') \right) },
\end{align*}

\noindent where $\xi_\delta(a) :=\nu(a)\left(1-e^{-\delta c(a)}\right)$ and $\zeta_\delta :=\sum_{a'\in\mathcal{A}}\nu(a')\,e^{-\delta c(a')}$.\\

Similarly, the tilted target marginal of the CPIP solution with parameter $\delta$ corresponds to:
\begin{align*}
    \nu^\star_\delta(a\given w) &= \frac{\nu(a)\sum_{a'\in\mathcal{A}}\pi(a'\given w)\,e^{-\delta c(a)\,\mathbb{I}(a'\neq a)}}{\sum_{a',a''\in\mathcal{A}}\pi(a'\given w)\,\nu(a'')\,e^{-\delta c(a'')\,\mathbb{I}(a'\neq a'')}} =  \frac{\nu(a)\left[\pi(a\given w) + \sum_{a'\neq a}\pi(a'\given w)\,e^{-\delta c(a)}\right]}{\sum_{a'\in\mathcal{A}} \pi(a'\given w)\left[\nu(a')+\sum_{a''\neq a'}\nu(a'')\,e^{-\delta c(a'')}  \right] },\\
    &=  \frac{\nu(a)\left[\pi(a\given w) + e^{-\delta c(a)}(1-\pi(a\given w))\right]}{\sum_{a'\in\mathcal{A}} \pi(a'\given w)\left[\nu(a')+\sum_{a''\in\mathcal{A}}\nu(a'')\,e^{-\delta c(a'')} -\nu(a')\,e^{-\delta c(a')} \right] },\\
    &=  \frac{\nu(a)\left[1 - (1-e^{-\delta c(a)})(1-\pi(a\given w))\right]}{\sum_{a'\in\mathcal{A}} \pi(a'\given w)\left(\zeta_\delta + \xi_\delta(a') \right) } =  \frac{\nu(a) - \xi_\delta(a)(1-\pi(a\given w)) }{\sum_{a'\in\mathcal{A}} \pi(a'\given w)\left(\zeta_\delta + \xi_\delta(a') \right) }.
\end{align*}

\LeftSectionStar{A4 -- Remark 3: Efficient influence functions}

Let $\nu$ and $c$ be given, $W$ to be a backdoor admissible set of pre-exposure covariates, $Q(w,a)=\mathbb{E}[Y\given W=w,A=a]$, and:
\begin{align*}
    \mathcal{S}_\delta[P] & \equiv \mu^S_\delta = \sum_{a\in\mathcal{A}}\mathbb{E}_W\left\{\pi^\star_\delta(a\given W)\, Q(W,a)\right\},\\ 
    \mathcal{T}_\delta[P] & \equiv \mu^s_\delta = \sum_{a\in\mathcal{A}}\mathbb{E}_W\left\{\nu^\star_\delta(a\given W)\, Q(W,a)\right\}, 
\end{align*}

Consider parametric submodel $P_\epsilon\in\mathfrak{P}$ indexed by a small fluctuation parameter $\epsilon\in\mathbb{R}$, and a point-mass contamination $O_i=(W_i,A_i,Y_i)\sim P^\circ $, such that, $P_\epsilon(O)=\epsilon\,\mathbb{I}({O}={O}_i)+(1-\epsilon)\,P^\circ ({O})$, where $P^\circ \in\mathfrak{P}$ is the true DGP distribution. Under some technical conditions involving  \emph{(i)} fully nonparametric or saturated model $\mathfrak{P}$,  \emph{(ii)} smoothness for the paths within the model, and  \emph{(iii)} boundedness of the outcome mean, the Gâteaux derivative and their variances, one has that $\mathcal{S}_\delta[P]$ and $\mathcal{T}_\delta[P]$ are pathwise differentiable at $P^\circ $.

The  \emph{uncentered} efficient influence function (EIF) of $\mathcal{S}_\delta[P]$ at $P^\circ $ evaluated at $O_i$ is given by $D_{\delta}^{{S}}(O_i) := \dv{\mathcal{S}_\delta[P_\epsilon]}{\epsilon}(O_i)\eval_{\epsilon=0}+\mathcal{S}_\delta[P^\circ ]$, and can be computed using the using the chain rule and gradient algebra for the Gâteaux derivative, as follows:
\begin{align*}
    D_{\delta}^{{S}}(O_i) &= \sum_{a\in\mathcal{A}}\frac{1}{H_\delta(W_i)^2}\left[H_\delta(W_i)\left(s_\delta(W_i,a)\,Q'(O_i,a)+s'_\delta(W_i,a)\,Q(W_i,a) \right)\right.\\
    &\qquad\left.-s_\delta(W_i,a)\,Q(W_i,a)\,H'_\delta(O_i)\right],\\
    &= \sum_{a\in\mathcal{A}}\left\{\frac{s_\delta(W_i,a)\,Q'(O_i,a)}{H_\delta(W_i)}+\frac{s'_\delta(W_i,a)\,Q(W_i,a)}{H_\delta(W_i)} - \frac{s_\delta(W_i,a)\,Q(W_i,a)\,H'_\delta(O_i)}{H_\delta(W_i)^2}\right\},
\end{align*}

with
\begin{align*}
    Q'(O_i,a) &= \frac{\mathbb{I}(a=A_i)}{\pi(a\given W_i)}\left[Y_i - Q(W_i,a) \right] + Q(W_i,a),  \\
    s_\delta(W_i,a) &= (\zeta_\delta +\xi_\delta(a))\,\pi(a\given W_i),\\
    s'_\delta(W_i,a) &= (\zeta_\delta +\xi_\delta(a))\, \mathbb{I}(a=A_i) - s_\delta(W_i,a),\\
    H_\delta(W_i) &= \sum_{a'\in\mathcal{A}} \left(\zeta_\delta + \xi_\delta(a') \right)\, \pi(a'\given W_i), \\
    H'_\delta(O_i) &=(\zeta_\delta +\xi_\delta(A_i)) - H_\delta(W_i).
\end{align*}

These expressions satisfy the following equivalences:
\begin{equation*}
    s_\delta(W_i,a')/H_\delta(W_i)  = \pi^\star_\delta(a'\given W_i)\quad\text{and}\quad
    (\zeta_\delta +\xi_\delta(a')) /H_\delta(W_i)  = \pi^\star_\delta(a'\given W_i) / \pi(a'\given W_i).
\end{equation*}

Therefore,
\vspace{-0.3cm}
\begin{align*}
    & D_{\delta}^{{S}}(O_i) =\sum_{a\in\mathcal{A}} \bigg\{\pi^\star_\delta(a\given W_i)\left[\frac{\mathbb{I}(a=A_i)}{\pi(a\given W_i)}\left[Y_i - Q(W_i,a) \right] + Q(W_i,a) \right]\\
    &\quad + \left[\pi^\star_\delta(a\given W_i)\frac{\mathbb{I}(a=A_i)}{\pi(a\given W_i)}-\pi^\star_\delta(a\given W_i)\right]\, Q(W_i,a) -\pi^\star_\delta(a\given W_i)\,Q(W_i,a)\left[\frac{\pi^\star_\delta(A_i\given W_i)}{\pi(A_i\given W_i)}-1 \right] \bigg\},\\
    &= \frac{\pi^\star_\delta(A_i\given W_i)}{\pi(A_i\given W_i)}\left[Y_i - Q(W_i,A_i) \right] + \sum_{a\in\mathcal{A}}\pi^\star_\delta(a\given W_i)\,Q(W_i,a) + \frac{\pi^\star_\delta(A_i\given W_i)}{\pi(A_i\given W_i)}\,Q(W_i,A_i)\\
    &\quad -  \frac{\pi^\star_\delta(A_i\given W_i)}{\pi(A_i\given W_i)} \sum_{a\in\mathcal{A}}\pi^\star_\delta(a\given W_i)\,Q(W_i,a),\\
    &= \underbrace{\frac{\pi^\star_\delta(A_i\given W_i)}{\pi(A_i\given W_i)}\left[Y_i - \sum_{a\in\mathcal{A}}\pi^\star_\delta(a\given W_i)\,Q(W_i,a) \right]}_{D_{\delta}^{{S},1}(O_i)} + \underbrace{\sum_{a\in\mathcal{A}}\pi^\star_\delta(a\given W_i)\,Q(W_i,a)}_{D_{\delta}^{{S},2}(O_i)}.
\end{align*}

Analogously, the  \emph{uncentered} EIF of $\mathcal{T}_\delta[P]$ at $P^\circ $  evaluated at point $O_i$ is given by $D_{\delta}^{{T}}(O_i) := \dv{\mathcal{T}_\delta[P_\epsilon]}{\epsilon}(O_i)\eval_{\epsilon=0}+\mathcal{T}_\delta[P^\circ ]$, and can be computed as:
\begin{equation*}
    D_{\delta}^{{S}}(O_i) = \sum_{a\in\mathcal{A}}\left\{\frac{t_\delta(W_i,a)\,Q'(O_i,a)}{H_\delta(W_i)}+\frac{t'_\delta(W_i,a)\,Q(W_i,a)}{H_\delta(W_i)} - \frac{t_\delta(W_i,a)\,Q(W_i,a)\,H'_\delta(O_i)}{H_\delta(W_i)^2}\right\},
\end{equation*}

\noindent where $t_\delta(W_i,a) = \nu(a) - \xi_\delta(a)(1-\pi(a\given W_i))$ and $t'_\delta(W_i,a) = \xi_\delta(a)\, \left[\mathbb{I}(a=A_i) - \pi(a\given W_i)\right]$.\\

These expressions satisfy the following equivalences:
\begin{align*}
    t_\delta(W_i,a')/H_\delta(W_i) & = \nu^\star_\delta(a'\given W_i),\\
    \xi_\delta(a')/H_\delta(W_i) & = \varrho_\delta(a')\, \pi^\star_\delta(a'\given W_i) / \pi(a'\given W_i), \text { with }\\
     \varrho_\delta(a') &= \xi_\delta(a') / (\zeta_\delta+\xi_\delta(a')).
\end{align*}

Therefore,
\begin{align*}
     & D_{\delta}^{{T}}(O_i) = \sum_{a\in\mathcal{A}} \bigg\{\nu^\star_\delta(a\given W_i)\left[\frac{\mathbb{I}(a=A_i)}{\pi(a\given W_i)}\left[Y_i - Q(W_i,a) \right] + Q(W_i,a) \right]\\ 
    &\quad + \left[ \varrho_\delta(a)\,\pi^\star_\delta(a\given W_i)\frac{\mathbb{I}(a=A_i)}{\pi(a\given W_i)}-\varrho_\delta(a)\,\pi^\star_\delta(a\given W_i)\right]\, Q(W_i,a)  -\nu^\star_\delta(a\given W_i)\,Q(W_i,a)\left[\frac{\pi^\star_\delta(A_i\given W_i)}{\pi(A_i\given W_i)}-1 \right] \bigg\},\\
    &=   \frac{\nu^\star_\delta(A_i\given W_i)}{\pi(A_i\given W_i)}\left[Y_i - Q(W_i,A_i) \right] + 2\sum_{a\in\mathcal{A}} \nu^\star_\delta(a\given W_i)\, Q(W_i,a) + \frac{\pi^\star_\delta(A_i\given W_i)}{\pi(A_i\given W_i)}\varrho_\delta(A_i)\,Q(W_i,A_i)  \\
    &\quad -\sum_{a\in\mathcal{A}} \pi^\star_\delta(a\given W_i)\,\varrho_\delta(a)\,Q(W_i,a) - \frac{\pi^\star_\delta(A_i\given W_i)}{\pi(A_i\given W_i)}\sum_{a\in\mathcal{A}} \nu^\star_\delta(a\given W_i)\, Q(W_i,a),\\
    &= \underbrace{\frac{\nu^\star_\delta(A_i\given W_i)}{\pi(A_i\given W_i)}\left[Y_i - Q(W_i,A_i) \right]}_{D_{\delta}^{{T},1}(O_i)} + \underbrace{\left[2 - \frac{\pi^\star_\delta(A_i\given W_i)}{\pi(A_i\given W_i)}\right]\sum_{a\in\mathcal{A}} \nu^\star_\delta(a\given W_i)\, Q(W_i,a)}_{D_{\delta}^{{T},2}(O_i)}\\
    &\quad + \underbrace{\frac{\pi^\star_\delta(A_i\given W_i)}{\pi(A_i\given W_i)}\varrho_\delta(A_i)\,Q(W_i,A_i)  -\sum_{a\in\mathcal{A}} \pi^\star_\delta(a\given W_i)\,\varrho_\delta(a)\,Q(W_i,a)}_{D_{\delta}^{{T},3}(O_i)}.
\end{align*}

\newpage
\LeftSectionStar{A5 -- Robustness to $Q$-misspecification and asymptotic linearity}

Let, $O=(W,A,Y)\sim P^\circ $, and $\pi^\circ(a\given w)=\mathbb{P}_{P^\circ }(A=a\given W=w)$, and $Q^\circ(z,a)=\mathbb{E}_{P^\circ }[Y\given Z=z,A=a]$, where $P^\circ \in\mathfrak{P}$ is the \emph{true} distribution of i.i.d observations $O_i$.

For any pair of measurable nuisance functions $(\widetilde\pi,\widetilde Q)$, let
$\widetilde\pi^\star_\delta(\cdot\given w)$ and $\widetilde\nu^\star_\delta(\cdot\given w)$ denote the tilted marginals obtained by applying the transformations
\eqref{eq:transf1}--\eqref{eq:transf2} with $\pi$ replaced by $\widetilde\pi$ (holding $\nu$ and $c$ fixed). Write $D^S_\delta(O;\widetilde\pi,\widetilde Q)$ and $D^T_\delta(O;\widetilde\pi,\widetilde Q)$ for the expressions in
\eqref{eqEIF1}--\eqref{eqEIF2} with $(\pi,Q)$ replaced by $(\widetilde\pi,\widetilde Q)$ and with
$\pi^\star_\delta,\nu^\star_\delta$ replaced by $\widetilde\pi^\star_\delta,\widetilde\nu^\star_\delta$.

\paragraph{(A5a) $Q$-robustness.} For any measurable $\widetilde Q$, if $\widetilde\pi=\pi^\circ$ then:
\[
\E_{P^\circ }\!\left[D^S_\delta(O;\pi^\circ,\widetilde Q)\right]=\mu^S_\delta,
\qquad
\E_{P^\circ }\!\left[D^T_\delta(O;\pi^\circ,\widetilde Q)\right]=\mu^T_\delta.
\]
Equivalently, the centered influence functions
$D^S_\delta(O;\pi^\circ,\widetilde Q)-\mu^S_\delta$ and
$D^T_\delta(O;\pi^\circ,\widetilde Q)-\mu^T_\delta$
have mean zero under $P^\circ $, even if $\widetilde Q\neq Q^\circ$.

\begin{proof}
We repeatedly use iterated expectation and the identity
$\E_{P^\circ }[\mathbb{I}(A=a)\given W]=\pi^\circ(a\given W)$.
Also, by the construction of $Z\subseteq W$ in the identification section,
$Q^\circ(Z,a)=\E_{P^\circ }[Y\given W,A=a]$ $P^\circ $-almost surely.

\paragraph{Case $\mu^S_\delta$.}
Condition on $W$ and abbreviate:
\[
m_{\widetilde Q}(W):=\sum_{a\in\mathcal{A}}\pi^\star_\delta(a\given W)\,\widetilde Q(Z,a),
\qquad
\omega_\delta(A,W):=\frac{\pi^\star_\delta(A\given W)}{\pi^\circ(A\given W)}.
\]
Under $\widetilde\pi=\pi^\circ$ we have $\widetilde\pi^\star_\delta=\pi^\star_\delta$, hence:
\[
D^S_\delta(O;\pi^\circ,\widetilde Q)=\omega_\delta(A,W)\,\{Y-m_{\widetilde Q}(W)\}+m_{\widetilde Q}(W).
\]
Taking $\E[\cdot\given W]$ one gets,
\begin{align*}
\E\!\left[D^S_\delta(O;\pi^\circ,\widetilde Q)\given W\right]
&=\E\!\left[\omega_\delta(A,W)Y\given W\right]
-\E\!\left[\omega_\delta(A,W)m_{\widetilde Q}(W)\given W\right]
+m_{\widetilde Q}(W)\\
&=\sum_{a\in\mathcal{A}}\pi^\circ(a\given W)\frac{\pi^\star_\delta(a\given W)}{\pi^\circ(a\given W)}\E[Y\given W,A=a]
-\;m_{\widetilde Q}(W)\sum_{a\in\mathcal{A}}\pi^\star_\delta(a\given W)+m_{\widetilde Q}(W)\\
&=\sum_{a\in\mathcal{A}}\pi^\star_\delta(a\given W)\,Q^\circ(Z,a),
\end{align*}
since $\sum_a \pi^\star_\delta(a\given W)=1$.
Taking $\E_W[\cdot]$ yields $\E[D^S_\delta(O;\pi^\circ,\widetilde Q)]=\mu^S_\delta$.

\paragraph{Case $\mu^T_\delta$.}
Again condition on $W$ and define:
\[
M_{\widetilde Q}(W):=\sum_{a\in\mathcal{A}}\nu^\star_\delta(a\given W)\,\widetilde Q(Z,a).
\]
When $\widetilde\pi=\pi^\circ$ we have $\widetilde\nu^\star_\delta=\nu^\star_\delta$ and $\widetilde\pi^\star_\delta=\pi^\star_\delta$, hence:
\begin{align*}
D^T_\delta(O;\pi^\circ,\widetilde Q)
&=\underbrace{\frac{\nu^\star_\delta(A\given W)}{\pi^\circ(A\given W)}\{Y-\widetilde Q(Z,A)\}}_{D^{T,1}}
+\underbrace{\Bigl[2-\omega_\delta(A,W)\Bigr]M_{\widetilde Q}(W)}_{D^{T,2}}\\
&\ +\underbrace{\omega_\delta(A,W)\,\varrho_\delta(A)\widetilde Q(Z,A)-\sum_{a\in\mathcal{A}}\pi^\star_\delta(a\given W)\varrho_\delta(a)\widetilde Q(Z,a)}_{D^{T,3}}.
\end{align*}
Each conditional expectation can be expressed as:
\begin{align*}
\E[D^{T,1}\given W]
&=\sum_{a\in\mathcal{A}}\nu^\star_\delta(a\given W)\Bigl(\E[Y\given W,A=a]-\widetilde Q(Z,a)\Bigr)
=\sum_{a\in\mathcal{A}}\nu^\star_\delta(a\given W)\Bigl(Q^\circ(Z,a)-\widetilde Q(Z,a)\Bigr),\\
\E[D^{T,2}\given W]
&=M_{\widetilde Q}(W)\Bigl(2-\E[\omega_\delta(A,W)\given W]\Bigr)
= M_{\widetilde Q}(W)\Bigl(2-\sum_{a\in\mathcal{A}}\pi^\circ(a\given W)\frac{\pi^\star_\delta(a\given W)}{\pi^\circ(a\given W)}\Bigr)
= M_{\widetilde Q}(W),\\
\E[D^{T,3}\given W]
&=\sum_{a\in\mathcal{A}}\pi^\circ(a\given W)\frac{\pi^\star_\delta(a\given W)}{\pi^\circ(a\given W)}\varrho_\delta(a)\widetilde Q(Z,a)
-\sum_{a\in\mathcal{A}}\pi^\star_\delta(a\given W)\varrho_\delta(a)\widetilde Q(Z,a)
=0.
\end{align*}
Therefore,
\[
\E[D^T_\delta(O;\pi^\circ,\widetilde Q)\given W]
=\sum_{a\in\mathcal{A}}\nu^\star_\delta(a\given W)\Bigl(Q^\circ(Z,a)-\widetilde Q(Z,a)\Bigr)
+\sum_{a\in\mathcal{A}}\nu^\star_\delta(a\given W)\widetilde Q(Z,a)
=\sum_{a\in\mathcal{A}}\nu^\star_\delta(a\given W)Q^\circ(Z,a).
\]
Taking $\E_W[\cdot]$ yields $\E[D^T_\delta(O;\pi^\circ,\widetilde Q)]=\mu^T_\delta$.
\end{proof}

\paragraph{(A5b) Asymptotic linearity of the one-step estimators}

To accommodate non-Donsker learners, we assume cross-fitting as described in the inference section:
let $\widehat D^{S}_\delta(O_i)$ and $\widehat D^{T}_\delta(O_i)$ denote foldwise EIF evaluations where
$(\widehat\pi,\widehat Q)$ are trained out-of-fold. We assume:

\begin{enumerate}
\item[(i)] \textit{Finite variance and bounded weights.}
$\E_{P^\circ }[Y^2]<\infty$, and there exists $C'<\infty$ such that
\[
\sup_{a\in\mathcal{A}}
\left|\frac{\pi^\star_\delta(a\given W)}{\pi^\circ(a\given W)}\right|
\;\le\; C',
\qquad
\sup_{a\in\mathcal{A}}
\left|\frac{\nu^\star_\delta(a\given W)}{\pi^\circ(a\given W)}\right|
\;\le\; C',
\qquad P^\circ \text{-almost surely.}
\]
\item[(ii)] \textit{Cross-fitting.} Each $\widehat D_\delta(\cdot)$ is evaluated on data independent of the sample used to fit
$(\widehat\pi,\widehat Q)$.
\item[(iii)] \textit{Nuisance rates.}
There exists a (possibly misspecified) limit regression $\bar Q$ with $\E[\bar Q(Z,A)^2]<\infty$ such that
$\|\widehat Q-\bar Q\|_{L^2(P^\circ )}=o_{P^\circ }(1)$ and
\[
\|\widehat\pi-\pi^\circ\|_{L^2(P^\circ )}=o_{P^\circ }(n^{-1/4}),
\qquad
\|\widehat Q-\bar Q\|_{L^2(P^\circ )}\,\|\widehat\pi-\pi^\circ\|_{L^2(P^\circ )}=o_{P^\circ }(n^{-1/2}).
\]
\item[(iv)] \textit{Stability of the policy maps.}
The maps $\pi\mapsto \pi^\star_\delta(\cdot\given W)$ and $\pi\mapsto \nu^\star_\delta(\cdot\given W)$ are Lipschitz in $L^2(P^\circ )$ in a neighborhood of $\pi^\circ$,
so that $\|\widehat\pi^\star_\delta-\pi^\star_\delta\|_{L^2(P^\circ )}\lesssim \|\widehat\pi-\pi^\circ\|_{L^2(P^\circ )}$ and
$\|\widehat\nu^\star_\delta-\nu^\star_\delta\|_{L^2(P^\circ )}\lesssim \|\widehat\pi-\pi^\circ\|_{L^2(P^\circ )}$. This holds in our setting because $\pi^\star_\delta(\cdot\given w)$ and $\nu^\star_\delta(\cdot\given w)$ are smooth rational functions of the finite-dimensional vector $\{\pi(a\given w)\}_{a\in\mathcal{A}}$ with denominators bounded away from $0$ by (i).
\end{enumerate}

Under this set of assumptions one has:
\[
\widehat\mu^S_\delta-\mu^S_\delta
=\frac{1}{n}\sum_{i=1}^n\Bigl(D^S_\delta(O_i;\pi^\circ,\bar Q)-\mu^S_\delta\Bigr)+o_{P^\circ }(n^{-1/2}),
\]
and
\[
\widehat\mu^T_\delta-\mu^T_\delta
=\frac{1}{n}\sum_{i=1}^n\Bigl(D^T_\delta(O_i;\pi^\circ,\bar Q)-\mu^T_\delta\Bigr)+o_{P^\circ }(n^{-1/2}).
\]
Consequently, $\widehat\mu^S_\delta$ and $\widehat\mu^T_\delta$ are asymptotically linear with influence functions
$D^S_\delta(\cdot;\pi^\circ,\bar Q)-\mu^S_\delta$ and $D^T_\delta(\cdot;\pi^\circ,\bar Q)-\mu^T_\delta$, respectively.
If $\bar Q=Q^\circ$ (consistent outcome regression), these coincide with the semiparametric efficiency bounds.

\begin{proof}
We prove the proof for $\widehat\mu^S_\delta$; the argument for $\widehat\mu^T_\delta$ is identical.
Let $P_n$ denote the empirical measure (implicitly cross-fitted), and write
$\widehat D^S_\delta(O):=D^S_\delta(O;\widehat\pi,\widehat Q)$ and
$D^{S,\circ}_\delta(O):=D^S_\delta(O;\pi^\circ,\bar Q)$.
By the rationale provided in the previous proof (5Aa) applied with $\widetilde Q=\bar Q$, one gets:
\[
P^\circ  D^{S,\circ}_\delta \;=\; \mu^S_\delta.
\]
Then one can produce the following decomposition:
\begin{align*}
\widehat\mu^S_\delta-\mu^S_\delta
&=P_n\widehat D^S_\delta-P^\circ D^{S,\circ}_\delta\\
&=(P_n-P^\circ )D^{S,\circ}_\delta
\;+\;P^\circ (\widehat D^S_\delta-D^{S,\circ}_\delta)
\;+\;(P_n-P^\circ )(\widehat D^S_\delta-D^{S,\circ}_\delta).
\end{align*}

\paragraph{Empirical-process remainder.}
By cross-fitting (ii), conditional on the training sample,
$\widehat D^S_\delta-D^{S,\circ}_\delta$ is fixed and evaluated on an independent sample.
Therefore,
\[
\E\!\left[\left.(P_n-P^\circ )(\widehat D^S_\delta-D^{S,\circ}_\delta)\right|\,\text{training}\right]=0,
\qquad
\operatorname{var}\!\left(\left.(P_n-P^\circ )(\widehat D^S_\delta-D^{S,\circ}_\delta)\right|\,\text{training}\right)
=\frac{1}{n}\|\widehat D^S_\delta-D^{S,\circ}_\delta\|_{L^2(P^\circ )}^2.
\]
Assumptions (iii)--(iv) imply $\|\widehat D^S_\delta-D^{S,\circ}_\delta\|_{L^2(P^\circ )}=o_{P^\circ }(1)$, hence
\[
(P_n-P^\circ )(\widehat D^S_\delta-D^{S,\circ}_\delta)=o_{P^\circ }(n^{-1/2}).
\]

\paragraph{Drift term is second order.}
A direct algebraic expansion of $P^\circ \widehat D^S_\delta-\mu^S_\delta$ yields a remainder of the form:
\[
P^\circ (\widehat D^S_\delta-D^{S,\circ}_\delta)
\;=\;
\mathcal{R}^S_\delta(\widehat\pi,\widehat Q;\pi^\circ,\bar Q),
\qquad
\text{with}\quad
\big|\mathcal{R}^S_\delta(\widehat\pi,\widehat Q;\pi^\circ,\bar Q)\big|
\;\lesssim\;
\|\widehat\pi-\pi^\circ\|_{L^2(P^\circ )}\,\|\widehat Q-\bar Q\|_{L^2(P^\circ )}
+\|\widehat\pi-\pi^\circ\|_{L^2(P^\circ )}^2.
\]
The bound follows from linearity of the EIF in $Q$ and the assumptions guaranteeing boundedness of the weighting ratios and Lipschitz stability of the induced policies. Assumption (iii) then gives $\mathcal{R}^S_\delta(\widehat\pi,\widehat Q;\pi^\circ,\bar Q)=o_{P^\circ }(n^{-1/2})$.

Putting all together one can show that:
\[
\widehat\mu^S_\delta-\mu^S_\delta
=(P_n-P^\circ )D^{S,\circ}_\delta+o_{P^\circ }(n^{-1/2})
=\frac{1}{n}\sum_{i=1}^n\Bigl(D^S_\delta(O_i;\pi^\circ,\bar Q)-\mu^S_\delta\Bigr)+o_{P^\circ }(n^{-1/2}),
\]
Since $D^S_\delta(O;\pi^\circ,\bar Q)$ has finite variance by assumption (i),
the classical Central Limit Theorem applies to $(P_n-P^\circ )D^{S,\circ}_\delta$, yielding asymptotic normality.
\end{proof}


\end{document}